\documentclass[twocolumn,aps,prl,showpacs,longbibliography,byrevtex]{revtex4-1}
\usepackage{epsf,graphicx,amssymb}
\usepackage[usenames]{color}
\usepackage{natbib}
\usepackage{float}
\usepackage{amsmath,amssymb}
\usepackage{color}

\begin{document}

\title{Tuning the resonant frequencies of a drop by a magnetic field\\
}
\author{Timoth{\'ee} Jamin}
\author{Yacine Djama}
\author{Jean-Claude Bacri}
\author{Eric Falcon}
\email[E-mail: ]{eric.falcon@univ-paris-diderot.fr}
\affiliation{Univ Paris Diderot, Sorbonne Paris Cit\'e, MSC, UMR 7057 CNRS, F-75 013 Paris, France}

\date{\today}
\begin{abstract}  

We report an experimental study of a magnetic liquid drop deposited on a superhydrophobic substrate and subjected to vertical vibrations in presence of a static magnetic field. It is well-known that a flattened drop of usual liquid displays oscillating lobes at its periphery when vibrated. By adding ferromagnetic nanoparticles to a water drop and varying the strength of the magnetic field, we are experimentally able to efficiently tune the resonant frequencies of the drop. By using conservation energy arguments, we show that the magnetic field contribution is equivalent to adding an effective negative surface tension to the drop. Our model is found in good agreement with the experiments with no fitting parameter.
\end{abstract}

%Magnetic fluids and ferrofluids, Drops and bubbles, Capillary waves

\maketitle

When a drop of liquid is dynamically driven by an external force, its free-surface generally displays an oscillating pattern at the drop resonant frequency.  At a fundamental level, the dynamical study of such an oscillating drop occurs in various domains at different scales. Indeed, liquid drop behavior is used to model stellar mass collisions or neutron-star oscillations in astrophysics \cite{Durisen86}, and nuclei in nuclear fission \cite{Amsden75,Zingman88}, or in cellular biology \cite{Brangwynne2011}. It is also involved in metrology for measuring viscosity of liquids \cite{Papoular97} or surface tension of molten materials \cite{Fraser71,Przyborowski95}. At a more practical level, it includes printing (ink drop generation), mixing of drop clouds, droplet manipulations in microfluidic, optofluidic, and pharmaceutical industry \cite{Oh12}. Since the pioneer works of Rayleigh \cite{Rayleigh79}, the resonant frequencies of a drop have been studied experimentally by means of different forcing mechanisms: water drop on a vibrating plate \cite{Yoshiyasu96,Okada06,Noblin09,Brunet11} or on a vibrating bath \cite{Dorbolo08}, in microgravity \cite{Apfel97}, by acoustic or air-flow levitation \cite{Shen10,Bouwhuis13}, by levitating it on its own vapor (Leidenfrost drops) \cite{Holter52,Snezhko08,Piroird13}, or by Lorentz force for a metal liquid drop \cite{Fautrelle99,Kocourek06}. A new challenge would be to accurately control and tune the free oscillations of a drop in a non-intrusive way. Notably, it would be of primary interest to be able to shift the resonant frequencies of the drop to avoid, for  instance, some annoying frequency bands in practical situations.

In this Letter, we study the dynamics of a flattened drop of magnetic fluid deposited on a superhydrophobic substrate vertically vibrated in presence of a weak magnetic field of tunable amplitude. The fluid used is a ferrofluid constituted of a stable suspension of nanometric magnetic particles diluted in a carrier liquid (water). Above a critical acceleration of vibrations, the drop undergoes a parametric instability leading to an azimuthal pattern around the drop that oscillates at half the forcing frequency. A star-shaped drop is then observed made of several oscillating lobes. We observe that the resonant frequencies of the drop depend on the strength of the field. We show quantitatively and theoretically that we are able to shift these frequencies by tuning the effective surface tension of the drop through the field. Indeed the contribution of the magnetic field is equivalent to the one of a negative surface tension. Including nanometric magnetic particles within a drop appears thus as a first step to be able to control and tune the free oscillations of a drop, such a magnetic fluid being known to shift the onsets of usual hydrodynamics instabilities (such as Kelvin-Helmholtz, Rayleigh-Taylor or Rayleigh-Plateau ones) \cite{Rosen,Cebers}. Finally, note that ferrofluid drop manipulation with a magnet is also an important task in microfluidics or lab on chip technology \cite{Bormashenko08,Sterr08,Nguyen10}, and in dynamic self-assembly phenomena \cite{Timonen13}.

\begin{figure}[t]
\begin{center}
\includegraphics[scale=0.5]{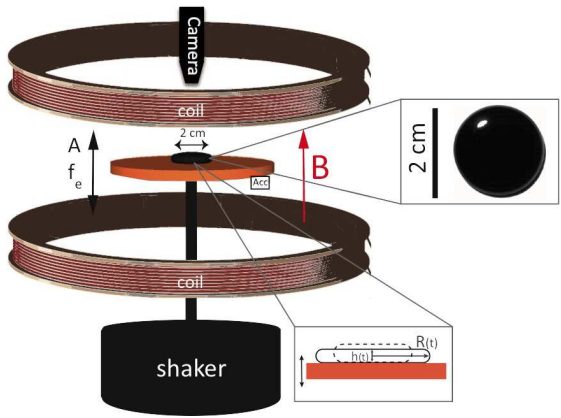}
\caption{(Color online) Experimental setup. Bottom inset shows a schematic of the puddle at two phases of the vibration. Top inset: top view of the ferrofluid puddle at rest.}
\label{fig02}
\end{center}
\end{figure}

The experimental set-up is shown in Fig.\ \ref{fig02}. A volume of ferrofluid ($V=1\,\mathrm{mL}$) is put on a plate. Due to gravity, the drop is flattened and looks like a puddle (radius, $R=9.7\,\mathrm{mm}$, and thickness, $h=3.4\,\mathrm{mm}$ - see insets of Fig.\ \ref{fig02}). The plate was coated with a spray \cite{UED}, to obtain a drop/substrate contact angle measured of roughly 165$^\circ$. Such superhydrophobic substrate is crucial to minimize the pinning force of the drop at the contact line, and thus to obtain much better reproducible experiments \cite{Celestini06,Chang13}. The plate is then subjected to vertical sinusoidal vibrations by means of an electromagnetic shaker. The frequency and amplitude of vibrations are in the range $0 \leq f_e \leq 52\,\mathrm{Hz}$ and $0\leq A \leq 3\,\mathrm{mm}$, respectively. An accelerometer is fixed below the plate to measure the normalized acceleration of vibrations $\Gamma=A\omega_e^2/g$ with $\omega_e=2\pi f_e$ and $g=9.81\,\mathrm{m.s^{-2}}$ the acceleration of gravity. One has $0 \leq \Gamma \leq 4$. The ferrofluid puddle on the plate is placed between two horizontal coils ($38\,\mathrm{cm}$ in mean diameter), generating a vertical static magnetic field $B$ in the range $0\leq B \leq 100\,\mathrm{G}$, 99\% homogeneous in the horizontal plane \cite{Browaeys99bis}. The drop oscillations are recorded by means of a high-speed camera (PhantomV10) located above the drop, with a 1000 frames per second sampling and a $1600 \times 1200$ pixels resolution. The ferrofluid used is an ionic aqueous suspension synthesized with 12.4\% by volume of maghemite particles ($\mathrm{Fe}_2\mathrm{O}_3$; $7\pm 0.3\,\mathrm{nm}$ in diameter) \cite{Talbot}. Nanoparticle diameter and magnetic field values are small enough to avoid sedimentation or agglomeration due to vibrations, gravity and magnetic fields \cite{Rosen}. Ferrofluid properties are: density, $\rho=1550\,\mathrm{kg.m^{-3}}$, surface tension, $\gamma=43\pm3\,\mathrm{mN.m^{-1}}$, magnetic susceptibility, $\chi(B)\approx \chi(B=0)=1$ \cite{chi}, magnetic saturation, $M_{sat}=36\times 10^{3}\,\mathrm{A.m^{-1}}$ and dynamic viscosity, $1.4\times 10^{-3}\,\mathrm{N.s.m^{-2}}$. Its capillary length is then $l_c=\sqrt{\gamma/(\rho g)}=1.7\,\mathrm{mm}$.

To accurately measure azimuthal drop oscillations, we use large flattened drops ($R \gg l_c$, see inset of Fig.\ \ref{fig02}). The Bond number, quantifying the ratio between gravity and capillary forces acting on the drop, is ${\rm Bo} \equiv (R/l_c)^2 \simeq 33$. The ratio between magnetic and capillary forces is quantified by the magnetic Bond number ${\rm Bo_m} \equiv B^2 R/ (\mu_0 \gamma$) with $R$ the puddle radius, $\mu_0=4\pi\times 10^{-7}\,\mathrm{H.m^{-1}}$ the magnetic permeability of the vacuum. For our range of $B$, one has ${\rm Bo_m} \leq 2$. Moreover, the magnetic effect is much smaller than gravity one (${\rm Bo_m}/{\rm Bo}\leq 0.06$) meaning that the strength of $B$ is weak enough to not deform the flattened region of the puddle at rest.

\begin{figure}[t]
\includegraphics[scale=0.55]{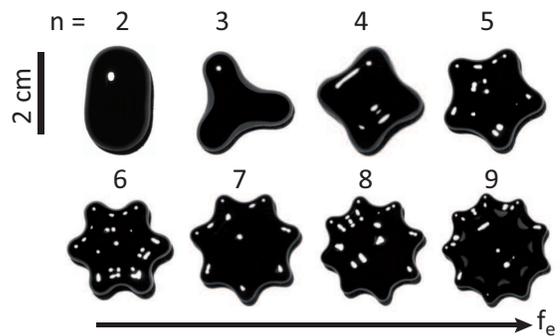} 
\quad\caption[]{Top view of the azimuthal pattern displayed around the drop as a function of $f_e$. Above a critical acceleration of vibration, an azimuthal pattern is observed normal to the vibration direction. When the forcing frequency $f_e$ is increased ($0 \leq f_e \leq 52\,\mathrm{Hz}$), the number $n$ of lobes, oscillating radially at $f_e$/2, increases from $n=2$ to 9 (from left to right and top to bottom). See \cite{Movie} for a movie. Magnetic field $B=0$. Volume $V=1\,\mathrm{mL}$}
\label{fig01}
\end{figure}

We first carry out experiments with no magnetic field ($B=0$). Figure \ref{fig01} shows the parametric instability of a ferrofluid puddle subjected to vertical sinusoidal vibrations. Above a critical acceleration of vibration $\Gamma_c$, an azimuthal pattern is observed in the horizontal plane at the drop periphery: lobes oscillate radially at half the forcing frequency, $f_e/2$. When $f_e$ is increased, the number $n$ of oscillating lobes (mode number) increases from $n=2$ to 9 as shown in Fig. \ref{fig01} and movie (see Supplemental Material \cite{Movie}).

The resonant frequencies $f_n$ of such an inviscid  drop are independent of the nature of the forcing and arise from an interplay between inertia and surface tension effects. In the limit $2R\gg h$, the flattened drop shape (see inset of Fig.\ \ref{fig02} is approximated by a cylindrical column of fluid. The radial amplitude of the lobes $a_n(t)$ is then governed by an harmonic oscillator equation of eigenfrequency $f_n$ \cite{Lamb,Rayleigh79}
\begin{equation}
\frac{\mathrm{d}^2a_n(t)}{\mathrm{d}t^2}+\omega_n^2 a_n(t)=0, \hspace{15pt} \omega^2_n= \frac{\gamma}{\rho R^3}n(n^2-1)
\label{fn}
\end{equation}
with $\omega_n=2\pi f_n$.

%Now, applying a vertical sinusoidal vibration of the {\color{red} substrate} forces parametrically the drop by modulating gravity {\color{red} $g \rightarrow g(t)=g\left[1+\Gamma \cos(\omega_e t)\right]$ with $\Gamma=A\omega_e^2/g$. This modulates the thickness $h$ of the puddle, that induces fluctuations of the drop radius $R(t)$ since $R=\sqrt{V/(\pi h)}$ by volume conservation. Consequently, a temporal modulation of gravity $g(t)$ induces a modulation of eigenfrequencies $\omega^2_n(t)\sim R(t)^{-3} \sim g(t)^{-3/4}$ using Eq. (\ref{fn}) and quasi-static approximation \cite{Yoshiyasu96}. For weak acceleration of vibrations $\Gamma \ll 1$, the parametric oscillator equation governing the amplitude of the lobes $a_n(t)$ then reads {\color{red} \cite{Yoshiyasu96}} 
%\begin{equation}
%\frac{d^2 a_n(t)}{dt^2}+\omega_n^2\left[1-\frac{3}{4}\Gamma\cos(\omega_e t)\right]a_n(t)=0
%\label{Mathieu}
%\end{equation}

Now, applying a vertical sinusoidal vibration of the substrate forces parametrically the drop by modulating gravity $g \rightarrow g(t)=g\left[1+\Gamma \cos(\omega_e t)\right]$ with $\Gamma=A\omega_e^2/g$. Besides, the thickness $h$ of the puddle is usually given \cite{Yoshiyasu96,Brunet11} by the quasi-static balance between gravity and capillary energies per unit volume ($2\gamma /h\simeq \rho gh/2$), i.e. $h\simeq 2l_c=2\sqrt{\gamma/(\rho g)}$. The hypothesis of constant volume of the puddle $V=\pi R^2 h$ then gives its radius $R=\sqrt{V/(\pi h)}$. Thus, a temporal modulation of gravity $g(t)$ induces those of the puddle thickness $h(t)\sim g(t)^{-1/2}$, radius $R(t)\sim h(t)^{-1/2} \sim g(t)^{1/4}$, and eigenfrequencies $\omega^2_n(t)\sim R(t)^{-3} \sim g(t)^{-3/4}$ using Eq. (\ref{fn}). 
For weak acceleration of vibrations $\Gamma \ll 1$, the parametric oscillator equation governing the amplitude of the lobes $a_n(t)$ then reads \cite{Yoshiyasu96}
\begin{equation}
\frac{d^2 a_n(t)}{dt^2}+\omega_n^2\left[1-\frac{3}{4}\Gamma\cos(\omega_e t)\right]a_n(t)=0
\label{Mathieu}
\end{equation}

Equation\ (\ref{Mathieu}) is the Mathieu's equation whose solutions are marginality curves (or instability ``tongues'') separating stable zones (no deformation of the drop) and unstable zones where azimuthal standing waves at the drop periphery oscillates at half the forcing frequency ($f_e/2$), near the resonant frequencies $f_n$ \cite{Mathieu}. This corresponds to the parametric instability shown in Fig.\ \ref{fig01}.

\begin{figure}
\begin{center}
\includegraphics[scale=0.45]{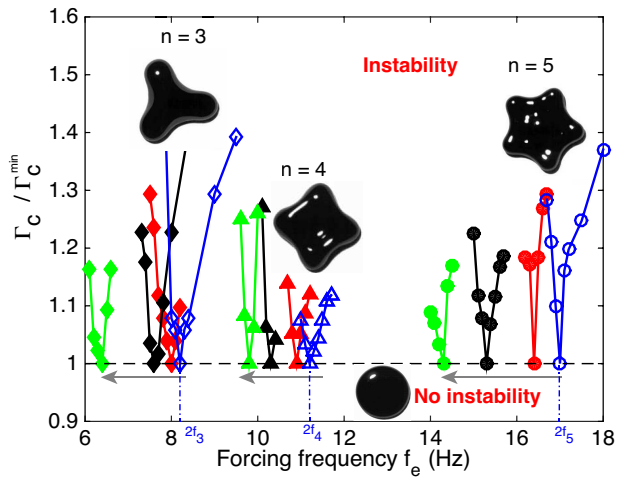}
\caption{(Color online) Phase diagram of normalized rescaled critical acceleration $\Gamma_c/\Gamma^{\rm min}_c$ vs. $f_e$. Curves are marginality curves separating stable and unstable zones for different mode numbers $n=3$ ($\lozenge$), 4 ($\vartriangle$) and 5 ($\circ$). No instability occurs for $\Gamma \leq \Gamma_{\rm c}$, whereas azimuthal pattern around the puddle occurs within tongues. Open symbols corresponds to $B=0$, full symbols to $B\neq 0$: $B=44$ (red, dark gray), 77 (black) and 99 G (green, light gray). When $B$ is increased (see arrows), the tongue are shifted towards a lower frequency, for each mode $n$.}
\label{fig03}
\end{center}
\end{figure}

These tongues of instability are displayed in Fig. \ref{fig03} in an acceleration/frequency phase space ($\Gamma_c / \Gamma^{\rm min}_c$ vs $f_e$)  for $B=0$ (open-symbols). Data were obtained experimentally for various $f_e$ by increasing $\Gamma$ until we observe lobes at critical acceleration $\Gamma_c$. $\Gamma^{\rm min}_c$ is the minimum critical acceleration of each tongue. Equation\ (\ref{Mathieu}) predicts that the minimum of each marginality curve occurs at twice the eigenmode $f_e= 2f_n$. Notice that experimental values for $f_n$ are smaller than values given by Eq.~(\ref{fn}). This difference ($<18\%$) may be explained by the fact that the quasi-static approximation yielding Eq.~\eqref{Mathieu} is valid only for low-frequency excitation. At higher frequency, the existence of axisymmetric modes shifts the minima of the marginality curves \cite{Noblin05,NoblinEPJE2004}. Indeed, we checked that this difference becomes negligible when we remove these axisymmetric modes by performing a control experiment with a superhydrophobic steady plate in contact with the drop top \cite{TheseJamin}. One can now wonder what is the effect of an applied vertical magnetic field $B$ on the drop dynamics.

\begin{figure}
\begin{center}
\includegraphics[scale=0.45]{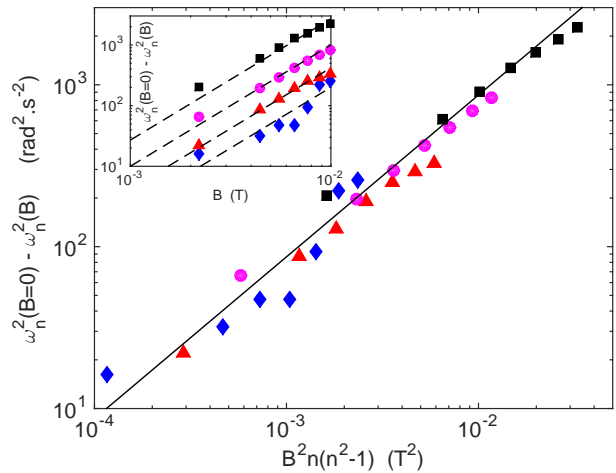} 
\caption{(Color online) Eigenfrequency shift $\omega_n^2(B=0) - \omega_n^2(B)$ as a function of $B^2n(n^2-1)$ for different modes $n=$3 ($\blacklozenge$), 4 ($\blacktriangle$), 5 ($\bullet$) and 7 ($\blacksquare$). Solid line is the prediction from the model of Eq.\ (\ref{theo}) with no fitting parameter. Inset: Unrescaled frequency shift vs.\! $B$. Dashed lines have a slope 2.}
\label{fig04}
\end{center}
\end{figure}

When $B$ is increased for a fixed $n$, we observe that the instability tongue is shifted towards a lower frequency (see arrows in Fig.\ \ref{fig03}). The minimum of this curve and thus the eigenfrequency $f_n(B)$ are found to decrease with $B$. For $n=5$, a relative shift of $f_n$ of 16\% is observed between extreme values of $B$ used. One defines the absolute shift of the eigenfrequency as $\Delta\Omega_n (B)\equiv \omega^2_n(B=0) - \omega_n^2(B)$, taking thus positive values. We plot in the inset of Fig.\ \ref{fig04}, the frequency shift $\Delta\Omega_n (B)$ as a function of $B$ for different $n$. We find that $\Delta\Omega_n (B) \sim B^2$ for our range of $B$ regardless of $n$. All data in the inset of Fig.\ \ref{fig04} are found to collapse on a single curve when plotting $\Delta\Omega_n (B)$ as a function of $B^2n^3$ (not shown). In order to compare with the model described below, $\Delta\Omega_n (B)$ is then displayed in the main Fig.\ \ref{fig04} as a function of $B^2n(n^2-1)$, noting that $n(n^2-1)\approx n^3$ for $n\geq 3$.

This frequency shift is not due to a geometrical effect mediated by $B$ such as a drop lengthening along the field direction \cite{Arkhipenko78,Bacri82,Cebers}. Indeed, using Eq.\ (\ref{fn}), a decrease of the puddle radius $R$ with $B$ (up to 6\% here) would lead to an increase of the resonant frequency, a situation opposite to our observations (see Fig.\ \ref{fig03}). Moreover, the results of Fig. 4 are found again when adding a steady plate above the drop. Finally, note that no clear dependence of the critical acceleration $\Gamma^{\rm min}_c$ on $B$ is observed. 

\begin{figure}[t]
\begin{center}
\includegraphics[scale=0.46]{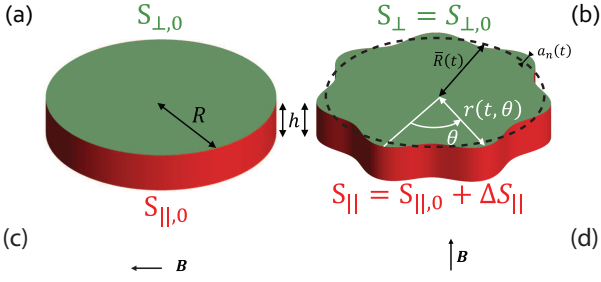}
\caption{(Color online) Schematic view of the puddle: (a) at rest, the puddle is considered as a cylinder of liquid, (b) when lobes appear, peripheral surface $S_\|$ increases as well as capillary energy. (c) and (d) illustrate magnetization within a ferrofluid film parallel or normal to $\mathbf{B}$.}
\label{puddle}
\end{center}
\end{figure}

An elegant way to understand the physical origin of the drop eigenfrequency shift with $B$ is to balance energies involved in this system. In a first step, we assume $B=~0$ and follow Rayleigh's model \cite{Rayleigh79}. The flattened drop shape is approximated by a cylinder of fluid (see Fig.\ \ref{puddle}a). We denote $S_\bot$ its surface area normal to gravity and magnetic field (i.e. top plus bottom areas) and $S_\|$ the peripheral surface area. At rest, they are denoted $S_{\bot,0}$ and $S_{\|,0}$. We consider small radial deformations of the peripheral surface of amplitude $a_n(t)\ll R$ around an instantaneous radius $\bar{R}(t)$. In polar coordinates, this reads $r(\theta,t)=\bar{R}(t)+a_n(t)\cos(n\theta)$ (see Fig.\ \ref{puddle}b). $h$ is assumed constant with time and thus also $S_\bot$ due to volume conservation. It is known, since Rayleigh \cite{Rayleigh79}, that the radial deformation $a_n$ induces an increase in $S_\|$ (see Figs.\ \ref{puddle}a-b) as
\begin{equation}
 \Delta S_\|(t) \equiv S_{\|}(t)-S_{\|,0}=\frac{\pi ha_n^2(t)(n^2-1)}{2R}
\label{S}
\end{equation} 
Thus, capillary energy increases by
\begin{equation}
\Delta E_c(t)=\gamma\Delta S_\|(t).
\label{E_c}
\end{equation} 
%The disc is assumed to be a slice  of a vertical column of fluid: when the cylinder is deformed 
Besides, kinetic energy of the liquid reads
\begin{equation}
E_k(t)=\left(\frac{\mathrm{d}a_n}{\mathrm{d}t}\right)^2\frac{\pi\rho hR^2}{2n}.
\label{E_k}
\end{equation} 
The conservation of energy $\mathrm{d}(E_k+E_c)/\mathrm{d}t=0$ then leads to Eq.\ (\ref{fn}) \cite{Rayleigh79}.

Let us now introduce the magnetic energy $E_m$ in the Rayleigh's model. For a linearly permeable ferrofluid of volume $V$ \cite{chi}, one has $E_m=-\mathbf{B}.\int_V \mathbf{M}dV/2$, with $\mathbf{B}$ the external magnetic field and $\mathbf{M}$ the local ferrofluid magnetization \cite{Rosen}. The determination of $\mathbf{M}$ needs to take into account the ferrofluid boundary conditions. For instance, for a plane parallel to $\mathbf{B}$, one has $E_m=-\chi VB^2/(2\mu_0)$ (see Fig.\ \ref{puddle}c) and for a plane normal to $\mathbf{B}$, $E_m=-\chi VB^2/[2\mu_0(1+\chi)]$ (see Fig.\ \ref{puddle}d). Thus, the magnetic energy of a ferrofluid layer is smaller when $\mathbf{B}$ is parallel rather than normal to its surface.

%Now, let us introduce the magnetic energy in this model. For a ferrofluid volume $V$, it may be written: $E_m=-\mathbf{B_0}/2.\int_V \mathbf{M}dV$ \cite{Rosen}, with $\mathbf{B_0}$ the external magnetic field and $\mathbf{M}$ the local ferrofluid magnetization. The determination of $\mathbf{M}$ needs to take into account boundary conditions due to demagnetizing effects. Let us introduce $\mathbf{H}$ through $\mathbf{B}=\mu_0(\mathbf{H}+\mathbf{M})=\mu_0(1+\chi)\mathbf{H}$. The continuity of the tangential component of $\mathbf{H}=\mathbf{B}/\mu$ (with $\mu=\mu_0$ outside the drop and $\mu=\mu_0(1+\chi)$ inside) induces $\mathbf{M}=\chi\mathbf{B}/\mu_0$ through an interface parallel to $\mathbf{B}$ (\textit{e.g.} $E_m=-\chi VB^2/2\mu_0$ for an infinite elongated cylinder parallel to $\mathbf{B}$. See Fig.\ \ref{puddle}.(c)). The continuity of the normal component of $\mathbf{B}$ induces $\mathbf{M}=\chi\mathbf{B}/\mu_0(1+\chi)$ through an interface normal to $\mathbf{B}$ (\textit{e.g.} $E_m=-\chi VB^2/2\mu_0(1+\chi)$ for an infinite flat cylinder normal to $\mathbf{B}$. See Fig.\ \ref{puddle}.(d)).

\begin{figure}[t]
\begin{center}
\includegraphics[scale=0.45]{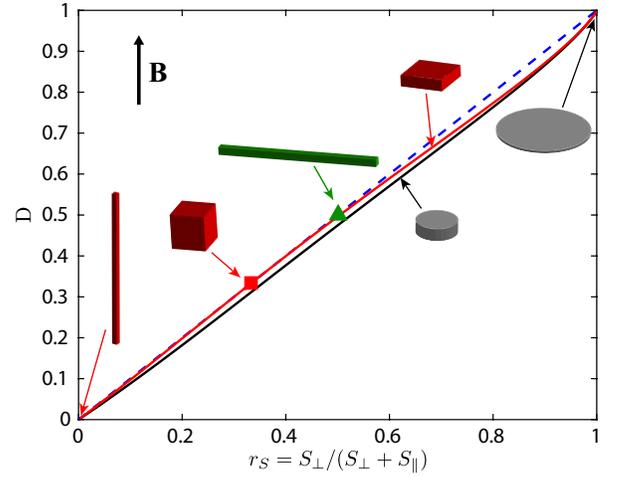}
\caption{(Color online) Theoretical demagnetizing factors $D$ vs. surface ratio $r_S$. Black line is computed from model in \cite{Arrott79} for a cylinder of axis aligned with $\mathbf{B}$; red line (light gray) is computed from \cite{Aharoni98} for a square rod aligned with $\mathbf{B}$; $\blacksquare$ is for a cube ($D=r_S=1/3$ \cite{Aharoni98}); $\blacktriangle$ is for an infinitely elongated square rod normal to $\mathbf{B}$ ($D=r_S=1/2$ \cite{Aharoni98})%; $+$ is for the cylinder approximating the drop of the present study
. Dashed line represents $D=r_S$ (slope 1).}
\label{D}
\end{center}
\end{figure}

For an arbitrary ferrofluid shape, $\mathbf{M}$ is nonuniform and an effective demagnetizing factor $D$ is usually defined, with $0\leq D\leq 1$ (depending on the shape), such that \cite{Rosen}
\begin{equation}
E_m=-\frac{\chi VB^2}{2\mu_0(1+\chi D)}.
\label{Em}
\end{equation}
For volumes bounded by surfaces either parallel or normal to $\mathbf{B}$, we can define the ratio between surface area normal to $\mathbf{B}$ and total surface area, $r_S \equiv S_{\bot }/(S_{\bot }+S_{\| })$. %=(1+S_{\| }/S_{\bot })^{-1}$
Using known theoretical values of $D$ for different geometries \citep{Arrott79,Aharoni98}, we show in Fig.\ \ref{D} that $D\approx r_S$ over the whole range of aspect ratios. This means that $E_m$ decreases when the aspect ratio favors surfaces parallel to $\mathbf{B}$, i.e. $D\rightarrow 0$ when $r_s\rightarrow 0$.  %In addition to infinite-cylinder examples considered in the above paragraph

We can then replace the demagnetizing factor $D$ by $r_S$ in Eq.~\eqref{Em}. The variation of $S_{\|}$ due to the presence of peripheral lobes induces a variation of $E_m$ through $r_S$. Noticing that $S_{\bot,0}=2\pi R^2$ and $S_{\|,0}=2\pi Rh$ at rest, a first-order Taylor expansion in $\Delta S_{\| }$ for small deformations ($a_n\ll R$) leads to
\begin{equation}
\Delta E_m(t)=\gamma_m \Delta S_\|(t)
\label{DEm}
\end{equation}
with
\begin{equation}
\gamma_m=-\frac{\chi^2 hB^2}{4\mu_0\left(1+\chi+h/R\right)^2},
\label{gammam}
\end{equation}
a quantity always negative.
%For the small variation $\Delta S_{\|}$, we have: $D(S_{\|})=D(S_{\|,0})+\Delta S_{\|}.D'(S_{\|,0})$ with $D_0$ the demagnetizing factor of the undeformed cylinder and a first-order Taylor expansion in $\Delta S_{\| }$ yields to $\Delta E_m=\chi^2VB^2\Delta S_{\bot}$.
%\begin{equation}
%\Delta E_m=-\frac{\chi^2 hB^2\Delta S_\|}{4\mu_0\left(1+\chi+h/R\right)^2}
%\end{equation}
Then, using $\Delta S_{\| }$ from Eq.\ (\ref{S}) and energies from Eqs~\eqref{E_c}, \eqref{E_k} and \eqref{DEm}, the conservation of energy $\mathrm{d}(E_k+E_c+E_m)/\mathrm{d}t=0$ finally leads to the resonant frequencies of the ferrofluid drop
\begin{equation}
\omega_n^2(B) = \frac{\gamma + \gamma_m(B^2)}{\rho R^3}n(n^2-1) {\rm \  \cdot}
\label{fnm}
\end{equation} 
Using Eq.\ (\ref{fn}) then leads to
\begin{equation}
\omega_n^2(B=0) - \omega_n^2(B) = - \frac{\gamma_m(B^2)}{\rho R^3}n(n^2-1){\rm \  \cdot}
\label{theo}
\end{equation}
The $B^2n(n^2-1)$ scaling is in good agreement with the one found experimentally (see Fig.\ \ref{fig04}) as well as for the theoretical prefactor without fitting parameter. 

Notice that Eq.\ (\ref{fnm}) includes the usual capillary contribution ($\gamma$) and a magnetic one ($\gamma_m$) that depends on $B$. The magnetic term thus plays the role of a negative surface tension ($\gamma_m<0$) that thus reduces the drop resonant frequencies. The magnetic field can be then used to tune the effective surface tension, $\gamma_{eff}\equiv \gamma + \gamma_m$, and thus $\omega_n$. For our ranges of $B$, using ferrofluid properties and geometry, one has $\gamma_m \in [-8.5,0]$ mN.m$^{-1}$, that is up to 20\% of $\gamma$. The analogy with surface tension arises from $\Delta E_m=\gamma_m(B^2)\Delta S_\|$. This means that an increase of the drop surface area parallel to $\mathbf{B}$ favors its magnetization and thus decreases $E_m$ since $\gamma_m<0$. Consequently, $B$ has a stabilizing effect on the lobes. Finally, note that a model of the dynamics of a ferrofluid drop confined between two plates \cite{Langer92,Jackson94} mentioned such a possible negative surface tension effect, but requires $\chi \ll 1$ and thus cannot apply here where $\chi=1$.

To conclude, we have studied the dynamics of parametric oscillations of a centimetric ferrofluid drop on a superhydrophobic plate subjected to vertical sinusoidal vibrations and a constant magnetic field. By adding ferromagnetic nanoparticles to a water drop, we are able to shift significantly its eigenfrequencies by tuning the magnetic field strength. Using energy conservation, we extend the Rayleigh's model and show that the resonant frequency shift is well captured by our model with no fitting parameter. We also show that the magnetic field acts as a negative surface tension, and is a way to tune the effective surface tension of the drop. Finally, the weakness of the field strength and the small size of ferromagnetic particles are favorable to miniaturization to plan to control the oscillations of centimeter-to-micro-scale drop in a new non-intrusive way for potential applications.

%Indeed, the oscillating lobes increase surface area at the periphery of the drop, and so increase  {\color{red} its} magnetization, {\color{red} a configuration more} energetically favorable  {\color{red} that then} decreases {\color{red} drop} resonant frequencies. 

%The weakness of the field strength, the nanometric sizes of the ferromagnetic particles and the non-intrusive control of the eigenmodes is favorable to miniaturization, and thus could have some interest for potential applications.

\begin{acknowledgments}
We thank D. Talbot for the ferrofluid synthesis, M. Berhanu, P. Brunet, M. Costalonga, and C. Laroche for fruitful discussions, and A. Lantheaume, Y. Le Goas and M.-A. Guedeau-Boudeville for technical help. T. J. was supported by the DGA-CNRS Ph.D. program. This work was partially financed by ANR Turbulon 12-BS04-0005.
\end{acknowledgments}
\bibliography{goutte}

\begin{thebibliography}{46}
\bibitem [{chi()}]{chi}%
  \BibitemOpen
  \href@noop {} {}\bibinfo {note} {For our range of values of $B$, the magnetic
  susceptibility of the ferrofluid $\chi(B)$ varies less than 3\%, thus one
  fixes $\chi(B)=\chi(B=0)$.}\BibitemShut {Stop}%
\end{thebibliography}
\end{document}